\documentclass[preprint]{aastex}

\slugcomment{Submitted to AJ}

\shorttitle{Analysis of three new planetary nebulae}
\shortauthors{Miranda, Pereira \& Guerrero}

\begin{document}


\title{Spectroscopic confirmation of the planetary nebula nature of PM\,1-242, 
PM\,1-318 and PM\,1-333 and morphological analysis of the nebulae}


\author{L.\,F. Miranda\altaffilmark{1}}
\affil{Instituto de Astrof\'{\i}sica de Andaluc\'{\i}a, CSIC, Ap. Correos 3004, 18080 Granada, 
Spain}
\email{lfm@iaa.es}

\author{C.\,B. Pereira\altaffilmark{2}}
\affil{Observat\'orio Nacional-MCT, Rua Jos\'e Cristino, 77. CEP 20921-400, 
S\~ao Cristov\~ao, Rio de Janeiro-RJ, Brasil}
\email{claudio@on.br}

\author{M.\,A. Guerrero\altaffilmark{1}}
\affil{Instituto de Astrof\'{\i}sica de Andaluc\'{\i}a, CSIC, Ap. Correos 3004, 18080 Granada, 
Spain}
\email{mar@iaa.es}


\begin{abstract}
We present intermediate resolution long-slit spectra and narrow-band H$\alpha$,  
[N\,{\sc ii}] and [O\,{\sc iii}] images of PM\,1-242, PM\,318 and PM\,1-333, three 
IRAS sources classified as possible planetary nebulae. The spectra show that the three objects 
are true planetary nebulae and allow us to study their physical properties; the images 
provide a detailed view of their morphology. PM\,1-242 is a medium-to-high-excitation 
(e.g., He\,{\sc ii}$\lambda$4686/H$\beta$ $\sim$ 0.4; [N\,{\sc ii}]$\lambda$6584/H$\alpha$ 
$\sim$ 0.3) planetary nebula with an elliptical shape containing [N\,{\sc ii}] 
enhanced point-symmetric arcs. An electron temperature [T$_{\rm e}$([S\,{\sc iii}])] of  
$\sim$ 10250\,K and an electron density [N$_{\rm e}$([S\,{\sc ii}])] of $\sim$ 2300\,cm$^{-3}$ 
are derived for PM\,1-242. Abundance calculations suggest a large helium abundance 
(He/H $\sim$ 0.29) in PM\,1-242. PM\,1-318 is a high-excitation 
(He\,{\sc ii}$\lambda$4686/H$\beta$ $\sim$ 1) planetary nebula 
with a ring-like inner shell containing two enhanced opposite regions, surrounded by a fainter 
round attached shell brighter in the light of [O\,{\sc iii}]. PM\,1-333 is an extended planetary 
nebula with a high-excitation (He\,{\sc ii}$\lambda$4686/H$\beta$ 
up to $\sim$ 0.9) patchy circular main body containing two low-excitation knotty arcs. 
A low N$_{\rm e}$([S\,{\sc ii}]) of $\sim$ 450\,cm$^{-3}$ and T$_{\rm e}$([O\,{\sc iii}]) 
of $\sim$ 15000\,K are derived for this nebula. Abundance calculations suggest that PM\,1-333 
is a type\,I planetary nebula. The lack of a sharp shell morphology, low electron 
density, and high-excitation strongly suggest that PM\,1-333 is an evolved planetary nebula. 
PM\,1-333 also shows two low-ionization polar structures whose morphology and emission properties 
are reminiscent of collimated outflows. We compare PM\,1-333 with other 
evolved planetary nebulae with collimated outflows and find that outflows among evolved planetary 
nebulae exhibit a large variety of properties, in accordance with these observed in 
younger planetary nebula.
\end{abstract}



\section{Introduction}

Planetary nebulae (PNe) represent the last evolutionary stage of 
low- and intermediate-mass (M $\leq$ 8 M$_{\odot}$) stars before they 
enter the white dwarf phase, forming from Asymptotic Giant Branch (AGB) stars after 
a short post-AGB phase (Bloecker 1995). Both the stellar and circumstellar properties 
dramatically change during the AGB to PN transition (e.g., Balick \& Frank 2002). A precise 
knowledge of the processes involved in this transition relies in the identification of 
a large number of these objects in order to establish possible connections and evolutionary 
paths from the AGB to the PN phase. Identification of 
candidate post-AGB stars and PNe has been successful using the IRAS [12]$-$[25] vs. 
[25]$-$[60] two colour diagram (Preite-Mart\'{\i}nez 1988; 
Garc\'{\i}a-Lario et al. 1997). 
In this diagram, these objects are located in a well defined region marked by 
two lines delimited by [12]$-$[25] $\geq$ 0.75 and [25]$-$[60] $\leq$ 1.13 
(see Garc\'{\i}a-Lario et al. 1997; Pereira \& Miranda 2007). 
However, this region overlaps with those of OH/IR stars, young stellar objects 
and  H\,{\sc ii} regions. Therefore, the IRAS two colour diagram does not provide an 
unambiguous identification of post-AGB stars and PNe.

Recently, we have started a spectroscopic program to establish the true 
nature of sources classified as possible post-AGB stars and PNe on the basis of 
their IRAS colours. We have already identified six new post-AGB stars, two H\,{\sc ii} regions, 
several objects of other different nature 
(Pereira \& Miranda 2007), and a new high density young PN, PM\,1-322, that could host 
a symbiotic star (Pereira \& Miranda 2005). As PM\,1-322 is located in a region of the 
IRAS two colour diagram that is common to post-AGB stars, PNe, OH/IR stars and young 
stellar objects, spectroscopy has been decisive to achieve an unambiguous classification.

In the framework of this spectroscopic program, we have also investigated the 
true nature of PM\,1-242, 
PM\,1-318 and PM\,1-333, three objects classified in SIMBAD as {\it possible planetary nebulae} 
mainly based in their IRAS colours. Table\,1 provides the names, coordinates and 
IRAS colours for these three objects. The three are located in a region of the IRAS two colour diagram 
in which PNe overlap with objects of different nature. In particular, it cannot be 
ruled out the possibility of PM\,1-242 and PM\,1-318 being compact H\,{\sc ii} regions 
(see Fig.\,1 in Pereira \& Miranda 2007). PM\,1-242 has 
been detected in the objective-prism survey of H$\alpha$ emission objects by 
Robertson \& Jordan (1989, their object 104). However, its detection in H$\alpha$ 
does not guarantee a PN nature. For PM\,1-318 there is no additional information 
available to investigate its true nature. Finally, in the case of 
PM\,1-333, radio continuum emission at 1.4 GHz and He{\sc ii}\,$\lambda$4686 line emission 
have been detected from the nebula (Condon et al. 1999), which does favor a 
PN classification. Nevertheless, 
a detailed analysis of the spectrum of PM\,1-333 has not been carried out as it is the case 
of PM\,1-242 and PM\,1-318 for which no spectra are available. In order to establish  
the nature of these objects, spectroscopic observations are a first crucial step. 

The morphology of these three objects has neither been analyzed. In the POSS, 
PM\,1-242 appears stellar, PM\,1-318 appears as a very small and tenuous 
nebulosity, and PM\,1-333 appears as a faint extended nebula. In none of the cases, 
details about the internal structure can be recognized in the POSS. Although morphology 
cannot be used as an indicator of the nature of a nebula, it is true that PNe and 
nebulae around post-AGB stars exhibit much more symmetric (or axisymmetric) shapes 
than H\,{\sc ii} regions (Manchado et al. 1996; Sahai \& Trauger 1998; Sahai et al. 2007). 
Furthermore, many PNe present small scale regions of high [N\,{\sc ii}]/H$\alpha$ ratios ($\geq$ 1) 
(e.g., Balick et al. 1994) that cannot be attained in young stellar objects and H\,{\sc ii} 
regions. Therefore, the presence of bright [N\,{\sc ii}] regions in a nebula can be 
indicative of a PN nature (e.g., Miranda et al. 
1998). Thus images obtained in [N\,{\sc ii}] and H$\alpha$ filters may be useful to assess the 
nature of a nebula.

In this paper, we present intermediate-resolution spectra and narrow-band optical images of 
PM\,1-242, PM\,1-318 and PM\,1-333. These observations have allowed us to confirm that they 
are true PNe, and to investigate their physical conditions and morphology.

\section{Observations}

\subsection{Spectroscopy}

Long-slit spectroscopic observations were obtained in 2005 August 3 and 5 using the Calar Alto 
Faint Object Spectrograph (CAFOS) at the 2.2\,m telescope on Calar Alto 
Observatory\footnote{The Centro Astron\'omico Hispano-Alem\'an at Calar Alto is 
operated jointly by the Max-Planck-Institut f\"ur Astronomie and the Instituto de 
Astrof\'{\i}sica de Andaluc\'{\i}a (CSIC).} (Almer\'{\i}a, Spain). 
The detector was a SITe CCD with 2048$\times$2048 pixels. The grisms B-100 and R-100 
were used to cover the spectral ranges 3200--6200 {\AA} and 5800--9600 {\AA}, respectively. 
In both spectral ranges, the dispersion was $\simeq$ 2 {\AA}\,pixel$^{-1}$. The spatial 
scale on the detector is 0.53\,arcsec\,pixel$^{-1}$. The slit 
width was 4$''$ and it was oriented North-South and centered in the objects; in the case 
of PM\,1-333, an additional red spectrum was obtained with the slit offset 27$''$ toward the West from the 
nebular center. The exposure time for both spectral ranges was 1800\,s in the 
cases of PM\,1-242 and PM\,1-318, 2700\,s for the spectra of the central region of 
PM\,1-333, and 3600\,s for the off-center spectrum of PM\,1-333. 
Spectrophometric standards from Oke (1974) and Massey et al. (1988) were observed 
for flux calibration. As the sky was clear but no photometric during the observations, the absolute 
fluxes must be used with caution.  The data were reduced using standard procedures within the 
IRAF\footnote{IRAF is distributed by the National Optical Astronomy Observatory, which is 
operated by the Association of Universities for Research in Astronomy (AURA) under cooperative 
agreement with the National Science Foundation.} package. 

\subsection{Imaging}

Narrow--band direct images of PM\,1-242, PM\,1-318 and PM\,1-333 were obtained in 2005 
August 3 and 2006 June 29 using ALFOSC at the Nordic Optical 
Telescope\footnote{Based on observations made with the Nordic Optical Telescope, 
operated on the island of La Palma jointly by Denmark, Finland, Iceland, Norway, 
and Sweden, in the Spanish Observatorio del Roque de los Muchachos of the Instituto 
de Astrof\'{\i}sica de Canarias.} on Roque de los 
Muchachos Observatory (La Palma, Spain). The detector was an E2V CCD 
with 2048$\times$2048 pixels and a plate scale of 0.19$''$\,pixel$^{-1}$. Images were
obtained through three narrow-band filters: H$\alpha$ ($\lambda$$_c$ = 6563 {\AA}, 
FWHM = 9 {\AA}), [N\,{\sc ii}] ($\lambda$$_c$ = 6584 {\AA}, FWHM = 9 {\AA}) and 
[O\,{\sc iii}] ($\lambda$$_c$ = 5007 {\AA}, FWHM = 30 {\AA}). Exposure time was 
1800\,s for each filter. Seeing was $\simeq$ 1$''$ during the observations. In 
the case of PM\,1-333, an additional image in the Johnson R filter was obtained 
during the acquisition of the spectra on 2005 August 3, with an exposure time of 300\,s 
and a seeing of $\simeq$ 1.2$''$. The images were reduced following standard 
procedures within the MIDAS package.

\section{Results}

Figure\,1 shows the blue and red spectra of PM\,1-242, PM\,1-318 and PM\,1-333. The spectra of PM\,1-242 and 
PM\,1-318 in Fig.\,1 have been obtained from the respective long-slit spectra by integrating  
over the total spatial extent of these two (relatively small) objects (see \S 3.1 and \S 3.2). 
In PM\,1-333, the spectra in Fig.\,1 correspond to two nebular regions of different emission properties 
extracted from the long-slit spectrum acquired with the slit going through the nebular 
center (see \S 3.3). All the spectra in Fig.\,1 are typical of ionized nebulae, showing emission lines of 
recombination H and He transitions, forbidden lines of various ionization states and weak 
continuum emission. Table\,2 presents the undereddened line intensities together with their 
Poissonian errors. The line intensities have been undereddened using the extinction law of Seaton (1979) and the 
logarithmic extinction coefficient c$_{\rm H\beta}$ (see Table\,2) derived from the observed 
H$\alpha$/H$\beta$ ratio assuming Case B recombination 
(T$_{\rm e}$ = 10$^4$\,K, N$_{\rm e}$ = 10$^4$\,cm$^{-3}$) for the theoretical Balmer line ratio 
of 2.85 (Brocklehurst 1971).  

The line intensities listed in Table\,2 show high He{\sc ii}\,$\lambda$4686/H$\beta$ and 
[O\,{\sc iii}]5007/H$\beta$ line ratios between 0.4 -- 1.1 and 5 --15, respectively. These 
high line ratios are not seen in H\,{\sc ii} regions and young stellar objects. On the contrary, they are 
typical of PNe. Therefore, the spectra of these three objects strongly point to a PN 
nature. Accordingly, we apply to these objects the PN\,G designation of PNe (Acker 
et al. 1992): PN\,G031.1+03.7, PN\,G074.5+02.3 and PN\,G100.4+04.6 for PM\,1-242, 
PM\,1-318 and PM\,1-333, respectively (see Table\,1). In the following, we will 
describe the spectra and images of each object in more detail. 

\subsection{PM\,1-242}

Figure\,2 shows the H$\alpha$, [N\,{\sc ii}] and [O\,{\sc iii}] images of PM\,1-242. 
The object presents an elliptical morphology with the major axis oriented almost along the 
North-South direction, and a size of $\simeq$ 8$''$$\times$5$''$ in [N\,{\sc ii}] and somewhat 
smaller in H$\alpha$ and [O\,{\sc iii}]. Two bright knots define the minor axis 
of the ellipse in the H$\alpha$ and [O\,{\sc iii}] images, with point-symmetric arcs 
emanating from these knots and tracing the ellipse. The point-symmetric intensity distribution 
is particularly marked in [N\,{\sc ii}]. At low intensity levels, 
the H$\alpha$ and [O\,{\sc iii}] images show protrusions along PA $\simeq$ 
50$^{\circ}$--230$^{\circ}$ and PA $\simeq$ 150$^{\circ}$ (see inset in the 
H$\alpha$ image of Fig.\,2). In addition, 
an isolated knot is observed in the three filters at 9.1$''$ from the center 
along PA $\simeq$ 225$^{\circ}$. Fig.\,2 also shows the [N\,{\sc ii}]/[O\,{\sc iii}] ratio 
map of PM\,1-242 constructed in order to analyze the excitation within the object. The lowest 
excitation is observed around the polar regions of the ellipse, in particular, in two 
point-symmetric arcs. The central region and bright knots along the minor axis 
show the highest excitation, while the isolated knot outside the main body of the 
nebula presents intermediate excitation. 

The spectrum of PM\,1-242 (Fig.\,1) shows relatively strong He\,{\sc ii} 
and [O\,{\sc iii}] emissions (He\,{\sc ii}$\lambda$4686/H$\beta$ $\simeq$ 0.43; 
[O\,{\sc iii}]$\lambda$5007/H$\beta$ $\simeq$ 15) and relatively weak  
[N\,{\sc ii}] and [S\,{\sc ii}] emissions ([N\,{\sc ii}]$\lambda$6584/H$\alpha$  
$\simeq$ 0.31; [S\,{\sc ii}]$\lambda$$\lambda$6716,6731/H$\alpha$ $\simeq$ 0.08). These line 
intensity ratios indicate a medium-to-high excitation PN. We note that variations of the 
line intensity ratios within the nebula should exist as indicated by the [N\,{\sc ii}]/[O\,{\sc iii}] ratio 
map (Fig\,2). In particular, the [N\,{\sc ii}]$\lambda$6584/H$\alpha$ ratio in the point-symmetric arcs 
is probably larger than 0.31.
From the [S\,{\sc ii}]$\lambda$6731/$\lambda$6716 line intensity ratio, we derive an electron 
density (N$_{\rm e}$) of 2300$\pm$660 cm$^{-3}$. The non detection of 
the [O\,{\sc iii}]$\lambda$4363 and [N\,{\sc ii}]$\lambda$5775 emission lines does not allow 
us to determine the electron temperature (T$_{\rm e}$) for these ions. Using the [S\,{\sc iii}] auroral and 
nebular lines detected in the spectrum, we obtain T$_{\rm e}$ = 10250$\pm$690\,K.

Chemical abundances in PM\,1-242 have been derived assuming the values of T$_{\rm e}$ and 
N$_{\rm e}$ quoted above. Ionic abundances are listed in Table\,3 and have been 
obtained using the task IONIC in the IRAF package. The errors in the ionic abundances account 
for the uncertainties in the line intensities and in the determination of the physical parameters. 
These errors are relatively high, particularly in the case of O$^+$ ($\sim$ 
60\%). Elemental abundances are listed in Table\,4 and have been obtained using the ionization correction 
factors (icf) of Kingsburgh \& Barlow (1994). Table\,4 also lists elemental abundances in other 
objects for comparison purposes. PM\,1-242 presents a high helium abundance 
(He/H $\simeq$ 0.29) that is well determined with an error of $\sim$ 10\%. This result 
suggests that PM\,1-242 is a type\,I PN although the N/O ratio of $\simeq$ 0.22 is unusually 
low for a type\,I PN (Peimbert 1990; see Table\,4). Note, however, that the errors in the ionic 
abundances of O$^+$ and N$^+$ (Table\,3) are large, and a N/O ratio of up to $\simeq$ 0.7  is 
consistent with the observations, making this value typical of type\,I PNe. A more accurate 
determination of the N/O ratio in this PN is most needed.

\subsection{PM\,1-318}

H$\alpha$ and [O\,{\sc iii}] images of PM\,1-318 are shown in Figure\,3 
together with an [O\,{\sc iii}]/H$\alpha$ image ratio derived to study the excitation within 
the object. The [N\,{\sc ii}] image is not shown because of the extreme weakness of 
the [N\,{\sc ii}] emission (see below). PM\,1-318 presents the morphology of a circular ring 
with a diameter of $\simeq$ 3.5$''$ and a thickness of $\simeq$ 1.5$''$, surrounded by an 
attached round shell with a diameter of $\simeq$ 10$''$ in [O\,{\sc iii}] and slightly smaller 
($\simeq$ 9$''$) in H$\alpha$. The ring contains two bright regions separated $\simeq$ 3.2$''$ 
along PA $\simeq$ 35$^{\circ}$. The [O\,{\sc iii}]/H$\alpha$ image ratio shows that the 
attached shell presents higher excitation than the inner nebular regions. 

The spectrum of PM\,1-318 is characterized by strong high-excitation lines and very faint 
low-excitation emissions (Fig.\,1 and Table\,2). In particular, He\,{\sc ii} emissions are 
strong (e.g., He\,{\sc ii}$\lambda$4686/H$\beta$ $\simeq$ 1.1). Other high-excitation emissions 
include [Ar\,{\sc v}] and [Cl\,{\sc iv}] lines. However, the [O\,{\sc iii}] emissions 
are moderate ([O\,{\sc iii}]$\lambda$5007/H$\beta$ $\simeq$ 
5.5) and the [O\,{\sc iii}]$\lambda$4363 line is not detected, suggesting relatively large 
amounts of O$^{+3}$ in the nebula. The [N\,{\sc ii}] emissions  
are very weak ([N\,{\sc ii}]$\lambda$6584/H$\alpha$ $\simeq$ 0.01) and  
[S\,{\sc ii}], [O\,{\sc i}] and [O\,{\sc ii}] emissions are absent. These results allow us to 
classify PM\,1-318 as a double shell high-excitation PN.

\subsection{PM\,1-333}

\subsubsection{Morphology}

Figure\,4 shows the H$\alpha$, [N\,{\sc ii}], [O\,{\sc iii}] and R images of PM\,1-333. 
The images show an extended nebula that is brighter in the light of [O\,{\sc iii}], 
somewhat fainter in the light of H$\alpha$, and that shows a very different morphology 
in the light of [N\,{\sc ii}]. All the images, in particular the R one, show a faint star 
located close to the geometrical center of the nebula, which may be 
considered as candidate to the central star of PM\,1-333. In the following, we will adopt 
this star as the origin for radial distances and PAs measurements.  We note that PM\,1-333 
is projected toward the young open cluster Trumpler\,37. 
This coincidence has allowed us to derive an R magnitude of $\simeq$ 17.8\,mag for the 
candidate central star by comparing it to the star Cl Trumpler\,37 Kun\,99 (Kun 
1986) which is registered in our R image. 

The [O\,{\sc iii}] image shows a nebula of $\simeq$ 105$''$$\times$50$''$ in size elongated 
along PA $\simeq$ 80$^{\circ}$. The main body of the nebula is almost circular with a 
size between 46$''$ and 50$''$.  No sharp inner or outer edge indicative of a shell 
structure is observed. The emission from this region is patchy and particularly bright in two 
knotty arcuate regions, one extending approximately North-South and one extending almost 
East-West. The morphology of the main body in H$\alpha$ is similar to that in [O\,{\sc iii}], although 
its extension is slightly smaller. In  [N\,{\sc ii}] only the two arcuate regions 
can be identified, while the diffuse emission is not detected. 

The [N\,{\sc ii}] image also reveals the presence of two knotty low-ionization structures 
(LIS) along PA $\simeq$ 70$^{\circ}$. The SW LIS is brighter than its NE counterpart and 
it extends closer (between 24$''$ and 37$''$) to the center of the nebula than the NE 
LIS (between 34$''$ and 43$''$).  Moreover, the morphology of both LIS is very different 
from each other: the SW LIS appears as a knotty 
filament elongated along PA $\simeq$ 205$^{\circ}$, while the NE LIS is composed of several 
individual knots. These LIS are associated to bright H$\alpha$ and [O\,{\sc iii}] extensions 
protruding from the central region. It is tempting to state that the [N\,{\sc ii}] 
bright knots are located at the tip of these high-excitation extensions, but a close 
inspection reveals that they are rather embedded within the high-excitation extensions. 
Furthermore, faint [O\,{\sc iii}] and H$\alpha$ emissions are 
detected beyond these LIS, although along PA $\simeq$ 80$^{\circ}$, 
slightly different from the PA $\simeq$ 70$^{\circ}$ orientation of the LIS.

The ionization structure of PM\,1-333 is illustrated in Figure\,5 where we present a 
colour composite picture constructed with the images in Fig.\,4. This colour picture reveals 
that [O\,{\sc iii}] dominates in the outer regions of the main body of the nebula while 
H$\alpha$ emission fills the central regions. Moreover, the bright [N\,{\sc ii}] emission 
of the arcuate structures appears embedded within regions of high excitation. We also note 
that some of the knots in the arcuate structures present radial tails that are better seen 
in the [N\,{\sc ii}] emission (red in Fig.\,5). In this figure, the LIS appear embedded 
within regions of high-excitation.

\subsubsection{Spectral analysis}

As already mentioned, spectra of PM\,1-333 were obtained through two different slit 
positions. The positions of these slits are overimposed on the R image of the nebula 
in Fig.\,4; a slit goes through the center of the nebula close to, but not including, the 
candidate central star, and the other slit covers the SW LIS. In the latter position, only 
a red spectrum was obtained. We will concentrate firstly on the spectrum of the main 
nebular body. 

The slit going through the center covers the low ionization arcuate structure toward the South, 
as well as the nebular regions toward the North where no [N\,{\sc ii}] is detected in the images. 
Consequently, we have extracted from the long-slit spectrum two one-dimensional spectra corresponding 
to the northern region, called PM\,1-333 (N), and to the southern arcuate structure, 
called PM\,1-333 (S). The spatial sizes of these regions are 21$''$ and 29$''$, respectively. 
These spectra are shown in Fig.\,1 (see also Table\,2). In PM\,1-333 (N), no 
low-ionization lines are observed, except for some faint [N\,{\sc ii}] emissions due to 
contamination from a low-ionization knot. In contrast, relatively strong [O\,{\sc i}], [O\,{\sc ii}], 
[S\,{\sc ii}] and [N\,{\sc ii}] 
emissions are observed in PM\,1-333 (S). The intensity of the He\,{\sc ii} lines, relative to H$\beta$, 
is higher in PM\,1-333 (N) than in PM\,1-333 (S), as it is the case of the high-excitation 
[Ar\,{\sc iv}]$\lambda$4740 line. He\,{\sc i} emissions are observed in PM\,1-333 (S) but 
not in PM\,1-333 (N). Other medium-to-high-excitation lines ([O\,{\sc iii}], [S\,{\sc iii}], 
[Ne\,{\sc iii}] and [Ar\,{\sc iii}]) are observed in the two regions but their intensity is relatively 
higher in PM\,1-333 (S) than in PM\,1-333 (N). In particular, the [O\,{\sc iii}]$\lambda$5007/H$\beta$ 
intensity ratio is high, $\simeq$ 12, in PM\,1-333 (S) but moderate, $\simeq$ 6, in PM\,1-333 (N). 
These results point out that the excitation level in PM\,1-333 (N) is high enough for 
O$^{+3}$, Ne$^{+3}$, Ar$^{+3}$ and S$^{+3}$ to be the prevalent ionization states. In PM\,1-333 (S), 
the arcuate filament preserves a lower excitation level and, therefore, larger amounts of O$^{+2}$, 
Ne$^{+2}$, Ar$^{+2}$ and S$^{+2}$.  

The auroral [N\,{\sc ii}]$\lambda$5755, [O\,{\sc iii}]$\lambda$4363 and [S\,{\sc iii}]$\lambda$6312 
emission lines detected in PM\,1-333 (S) allow us to derive values for the electron temperature: 
T$_{\rm e}$([N\,{\sc ii}]) = 17600$\pm$620\,K, T$_{\rm e}$([O\,{\sc iii}]) = 15300$\pm$225\,K and 
T$_{\rm e}$([S\,{\sc iii}]) = 22300$\pm$1900\,K. A rather 
low N$_{\rm e}$ of 445$\pm$45\,cm$^{-3}$ is derived from the [S\,{\sc ii}] line intensity ratio. 
This value corresponds to the electron density of the low ionization arcuate structure, not 
to the surrounding higher excitation material in which the electron density is probably lower.

Chemical abundances for PM\,1-333 (S) have been derived using the task IONIC in IRAF and  
the icf of Kingsburg \& Barlow (1994), assuming N$_{\rm e}$ = 445$\pm$45 cm$^{-3}$ and 
T$_{\rm e}$ = 15300$\pm$225\,K. The derived ionic and elemental abundances are 
listed in Table\,3 and Table\,4, respectively. Errors in the ionic abundances are 
$\leq$ 10\%. In consequence, errors in elemental abundances are estimated to be 
20\% -- 40\%. The relatively high helium abundance (He/H $\simeq$ 0.13) and high N/O ratio 
($\simeq$ 0.6) point out to a type\,I PN (Peimbert 1990; 
Stanghellini et al. 2006), although the total abundances of N, O, Ne, Ar and S appear lower than 
the average in PNe (see Kingsburgh \& Barlow 1994). 

Finally, the red spectrum obtained through the SW LIS (see Fig.\,4) is shown in Figure\,6. The 
undereddened line intensities are listed in Table\,2 and have been derived assuming that the 
logarithmic extinction coefficient in the SW LIS is identical to this in PM\,1-333 (S) (c$_{H\beta}$ = 
1.19, see Table\,2). We note that 
the derived line intensity ratios do not critically depend on the assumed value of 
c$_{H\beta}$. The spectrum of the SW LIS is characterized by very strong low-excitation emissions. 
The intensities of the [N\,{\sc ii}], [O\,{\sc i}], [O\,{\sc ii}], and [S\,{\sc ii}] emissions 
relative to H$\alpha$ are higher by a factor $\simeq$ 3 in the SW LIS than 
in the arcuate structure. High-excitation emissions of [Ne\,{\sc iii}] and 
[Ar\,{\sc iii}] are also observed in the SW LIS with a relative intensity comparable to or 
higher than those observed in the main nebular body. These emissions probably arise in the high-excitation 
material in which the SW LIS is embedded. The electron density in the SW LIS, derived from the 
[S\,{\sc ii}] line intensity ratio, is $\simeq$ 55\,cm$^{-3}$, at the low density limit 
of the [S\,{\sc ii}] intensity ratio and lower than the electron density derived above for the 
arcuate structure. 
The electron temperature, derived from the [S\,{\sc iii}] line intensity ratio, is 13500$\pm$1600\,K, 
lower than the T$_{\rm e}$([S\,{\sc iii}]) derived in the main nebular body.

\section{Discussion and conclusions}

We have presented intermediate-resolution long-slit spectra and narrow-band 
direct images of three objects classified in SIMBAD as possible PNe: PM\,1-242, PM\,1-318 and 
PM\,1-333. For PM\,1-242 and PM\,1-318, evidence for their real nature was lacking. 
For PM\,1-333, the detection of radio continuum emission and He\,{\sc ii} line emission 
suggested a PN nature (Condon et al. 1999) but a detailed study of the object was not 
available.

The spectral properties of PM\,1-242, PM\,1-318 and PM\,1-333 described in \S3 
confirm that they are indeed PNe. PM\,1-242 presents both high- and low-ionization 
line emissions typical of a medium-to-high-excitation PN. PM\,1-318 exhibits a high-excitation 
spectrum with strong He\,{\sc ii} and [Ar\,{\sc v}] line emissions and extremely 
weak or absent low-ionization emissions. Finally, PM\,1-333 presents regions with 
different excitation: high-excitation in the main nebular body and low-excitation in 
two polar structures and two arcs embedded in the main nebular body. 

PM\,1-242 exhibits an elliptical morphology with two point-symmetric, low ionization 
[N\,{\sc ii}] enhanced arcs tracing the ellipse. The physical structure 
of PM\,1-242 could be an ellipsoid, with the major axis along the North-South 
direction, containing two bright point-symmetric arcs and a bright, high excitation 
equatorial region. In this case, the physical structure of PM\,1-242 would be 
similar to this of Cn\,3-1 as shown by studies of its internal kinematics 
(Miranda et al. 1997). Alternatively, the faint protrusions observed in the 
H$\alpha$ and [O\,{\sc iii}] images suggest that PM\,1-242 could be a bipolar 
PN with its bipolar axis along the East-West direction. The bright ellipse 
can then be interpreted as a tilted equatorial ring while the faint protrusions (and the 
isolated knot) would trace faint bipolar extensions. The ring-like structure of 
PM\,1-242 would be then similar to this of, e.g., He\,2-428 (Manchado et al. 1996), 
IC\,2149 (V\'azquez et al. 2002) or Me\,1-1 (Pereira et al. 2008). These PNe present the 
same basic structure consisiting in a bright equatorial ring accompanied 
by faint bipolar extensions, the difference among them being the orientation of the 
ring axis (or bipolar axis) with respect to the observer. A study of the internal 
kinematics of PM\,1-242 would certainly allow us to determine the real structure of PM\,1-242. 

PM\,1-318 consists of a ring-like inner shell containing two enhanced opposite regions, surrounded 
by a fainter circular attached shell. The attached shell presents higher excitation than the inner 
one. The morphology of PM\,1-318 is similar to this of, e.g., NGC\,6826, NGC\,3242 
(Balick et al. 1998) and M\,2-2 (Manchado et al. 1996). The circular morphology of the 
inner shell suggests that PM\,1-318 is a round PN. Nevertheless, it cannot be ruled 
out that the inner shell is elliptical with the major axis tilted with respect to the 
line of sight. 

PM\,1-333 is probably an evolved PN, as strongly pointed out by the data presented in 
this paper. The absence of a 
sharp shell morphology suggests that the fast wind from the central star has ceased 
and that the nebular material is backfilling the central cavity. The evolved 
nature is further confirmed by the high-excitation of the nebula, suggesting a high 
temperature for the central star, and by the low nebular electron density. In this 
context, the bright low ionization arcs may represent relics of presumably dense, low ionization 
structures that were previously present in the nebula. The arcuate morphology of these 
low ionization structures suggests that they may have been tracing a ring or 
filamentary structures as those seen in NGC\,6543 (Miranda \& Solf 1992; Balick \& 
Hajian 2004). The LIS present the properties observed in FLIERs of PNe 
(Balick et al. 1998). The morphological differences between the SW LIS and the NE LIS compare 
to these observed in NGC\,7009 in which the FLIERs exhibit different morphologies and contain 
knots oriented at different directions with respect to the central star (Balick et al. 1998). 
The LIS strongly suggest that collimated outflows are present in PM\,1-333. Moreover, 
the protrusions connecting the LIS to the main nebular body point out that the 
collimated outflows have interacted with nebular material in the main nebular body 
deforming an original round shell. An analysis of the internal kinematics of PM\,1-333 
would be very valuable in order to study the formation processes of this interesting PN.

Although collimated outflows are usually observed in PNe, 
there are only a few cases of evolved PNe in which these structures have been detected. 
In this respect, PM\,1-333 can be compared to NGC\,1360 (Goldman et al. 2004), Wray\,17-4 and K\,1-2 
(Corradi et al. 1999). These four PNe present a patchy main body with no sharp shell morphology, 
indicative of evolved PNe, and collimated structures characterized by enhanced [N\,{\sc ii}] emission. 
However, the four cases exhibit noticeable differences. The main shell is approximately circular 
in Wray\,17-4 and PM\,1-333 and elliptical in NGC\,1360 and K\,1-2. The collimated structures are 
embedded in the main shell of Wray\,17-4 and K\,1-2, outside the main shell 
but connected to it in PM\,1-333, whereas in NGC\,1360 the collimated structures are outside the main 
shell but not connected to it. The orientation of the outflows relative to that of the main shell is 
also different in each object. In  NGC\,1360, the outflows are oriented 
along the major axis of the elliptical shell, in K\,1-2 they are oriented almost 
perpendicular the main axis of the shell, while in PM\,1-333 and Wray\,17-4 
no orientation can be established for their circular shells. As for the kinematics, 
the outflows in Wray\,17-4 and  K\,1-2 seem to share the kinematics of the shell while in NGC\,1360 
they present radial velocities larger than the shell. These results show that collimated 
outflows in evolved PNe present the same varied phenomenology as those observed in younger 
PNe (see, e.g., Sahai \& Trauger 1998, Miranda et al. 1999, Guerrero et al. 2008, and references 
therein). This variety is probably related to the different physical processes involved 
in the formation of PNe.

\acknowledgments

We thank an anonymous referee for comments that have improved the presentation and discusion 
of the data. We are grateful to the staff of the NOT and Calar Alto observatories 
for their help during the observations. This work has been supported partially 
by grants AYA2005-01495 of the Spanish MEC and AYA2008-01934 of the Spanish MICINN (co-funded by FEDER 
funds), and by grant FGM-1747 of the Junta de Andaluc\'{\i}a. This research has made use of 
the SIMBAD database, operated at CDS, Strasbourg, France.

{}

\clearpage

\clearpage

\begin{figure} 
\figurenum{1} 
\epsscale{0.95} 
\plotone{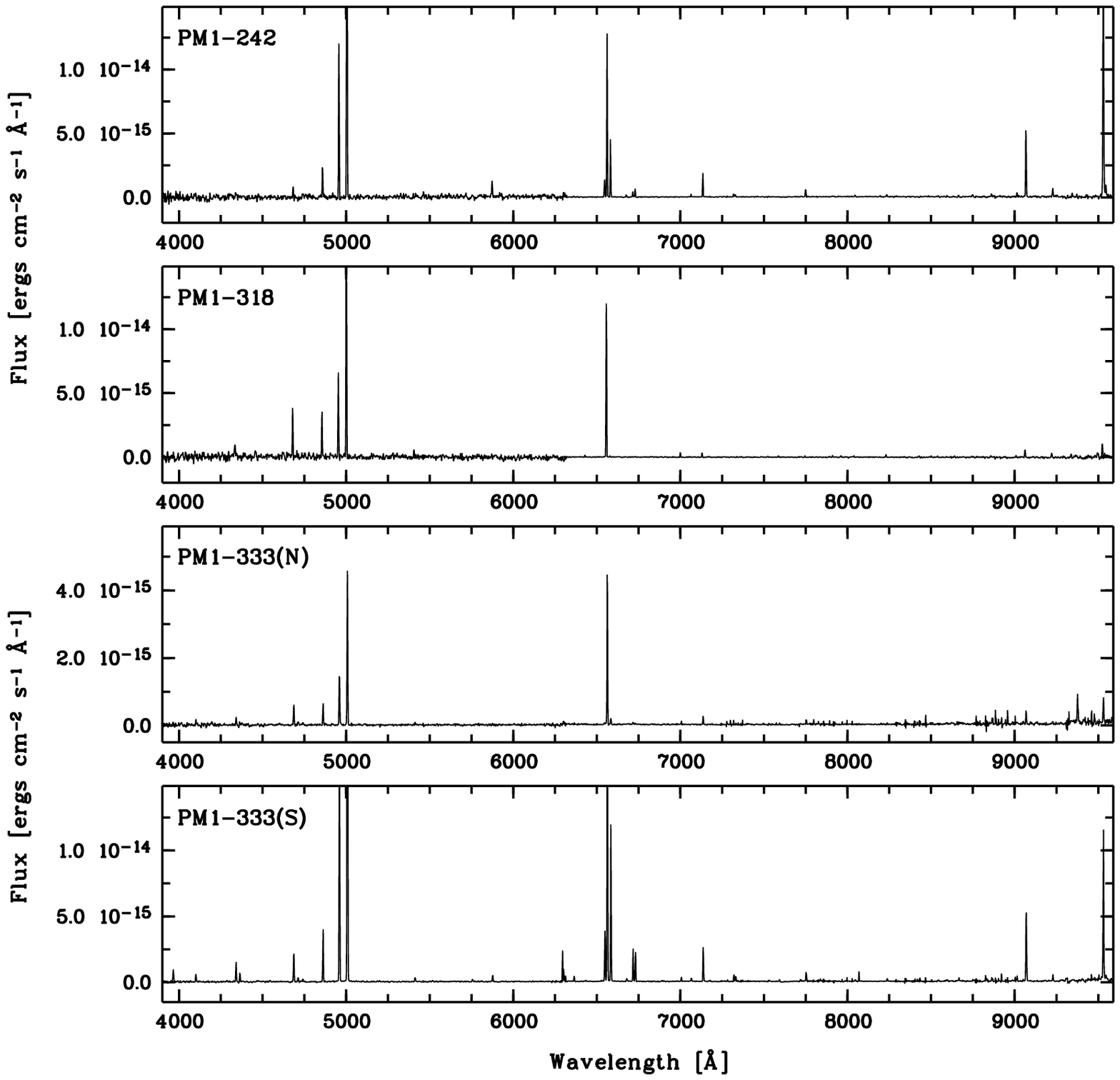} 
\caption{Blue and red 2.2\,m CAHA CAFOS spectra of 
PM\,1-242, PM\,1-318 and PM\,1-333. In the case of PM\,1-333, the spectra of 
two different regions (North and South) within the object are shown (see text).} 
\end{figure} 

\clearpage

\begin{figure} 
\figurenum{2} 
\epsscale{0.85} 
\plotone{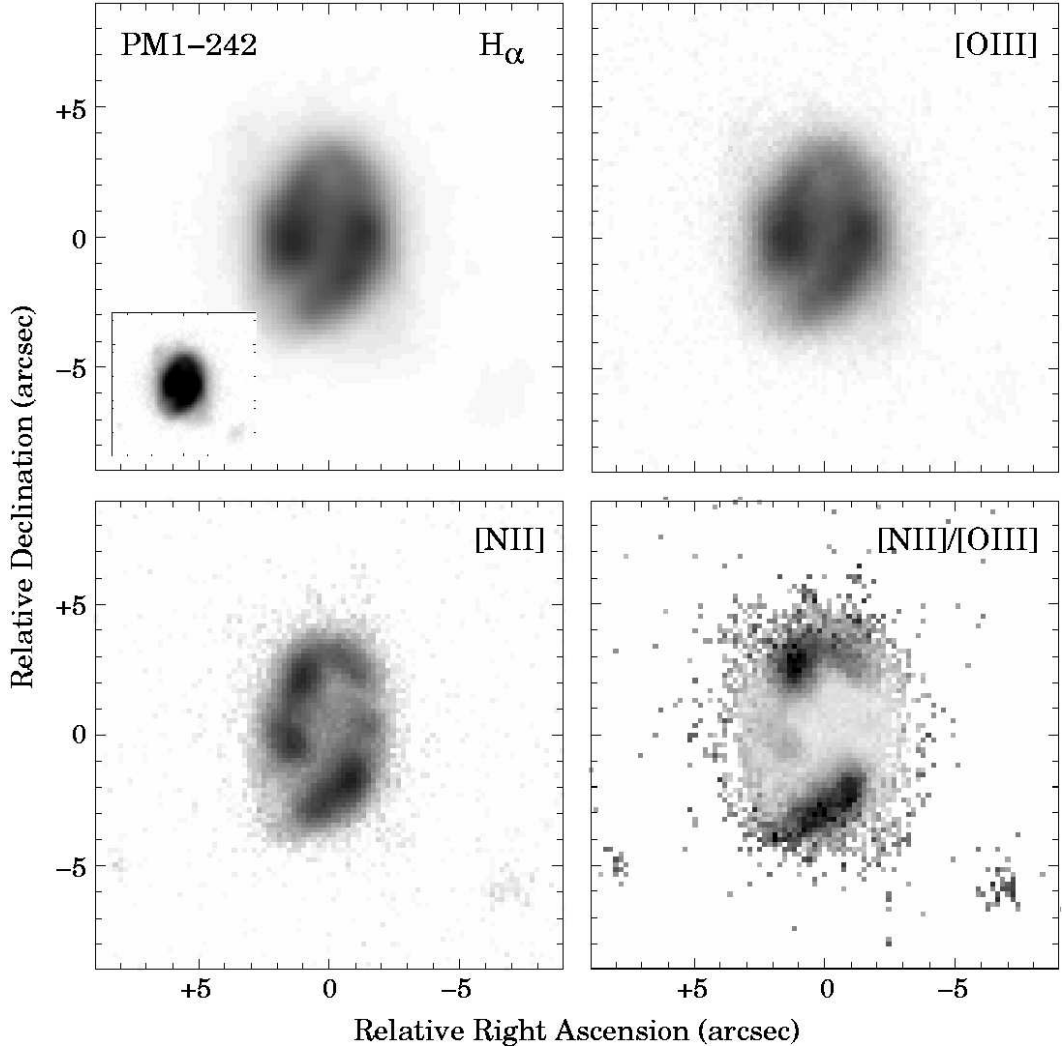} 
\caption{Grey-scale representations of the 
H$\alpha$, [N\,{\sc ii}] and [O\,{\sc iii}] images, and [N\,{\sc ii}]/[O\,{\sc iii}] image ratio 
of PM\,1-242. Grey levels are logarithmic except in the [N\,{\sc ii}]/[O\,{\sc iii}] image ratio 
where they are linear (black represents high values of the ratio). 
The H$\alpha$ image is shown in 
the inset with a different grey scale in order to highlight the faint regions of the object. 
(0,0) corresponds to the geometrical 
center of the two bright knots oriented east-west in the H$\alpha$ and [O\,{\sc iii}] images.} 
\end{figure}

\clearpage

\begin{figure} 
\figurenum{3} 
\epsscale{1.0} 
\plotone{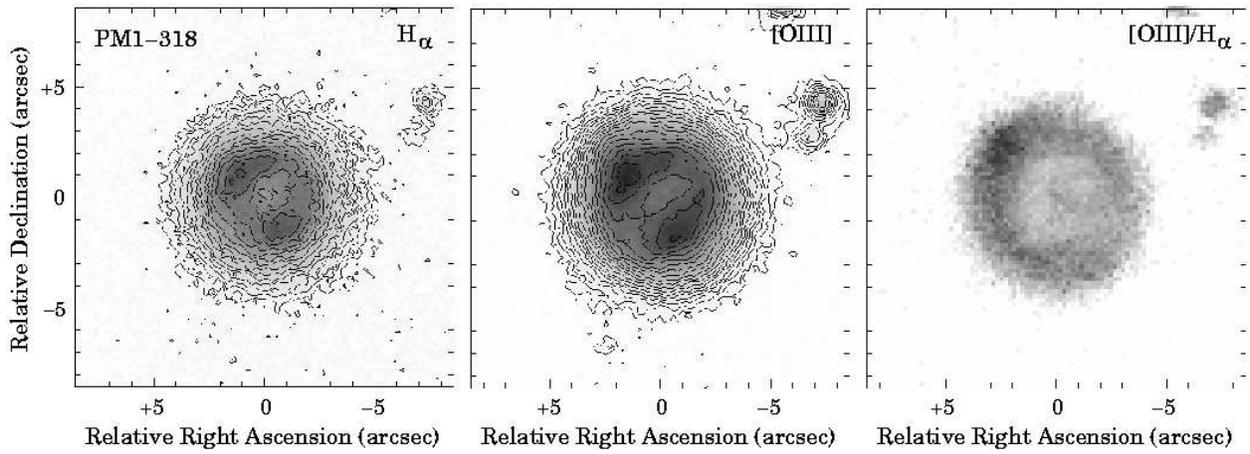} 
\caption{ Grey-scale and contour representations of the 
H$\alpha$ and [O\,{\sc iii}] images, and [O\,{\sc iii}]/H$\alpha$ image ratio of PM\,1-318. The 
grey levels are linear. Contours are arbitrary and have been chosen to highlight 
the structure of the nebula. In the [O\,{\sc iii}]/H$\alpha$ image ratio, black represents 
high values of the ratio. (0,0) corresponds to the position of the central minimum 
in the H$\alpha$ image.} 
\end{figure} 

\clearpage

\begin{figure} 
\figurenum{4} 
\epsscale{0.95} 
\plotone{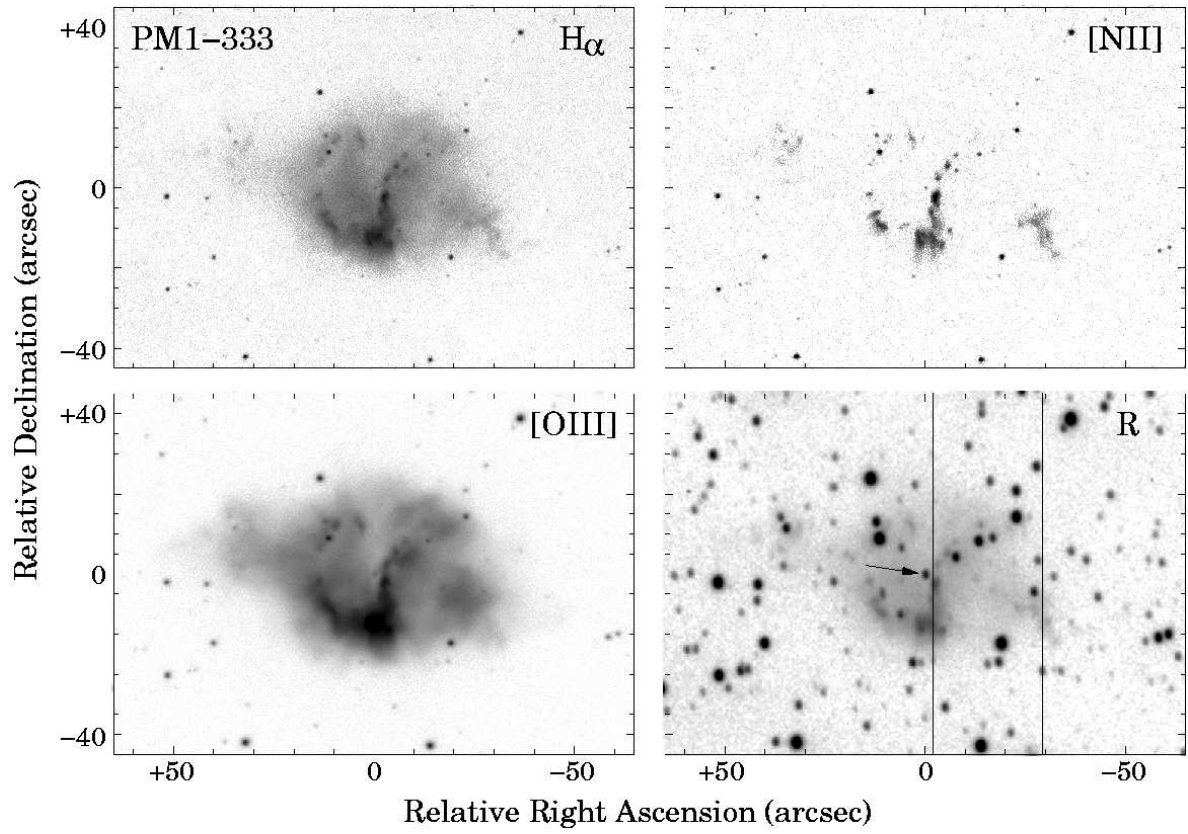} 
\caption{ Grey-scale representations of the 
H$\alpha$, [N\,{\sc ii}], [O\,{\sc iii}] and R images of PM\,1-333. Grey levels are 
logarithmic. (0,0) corresponds to the 
position of the candidate central star of the nebula, indicated in the R image by an arrow. The 
two vertical lines in the R image indicate the two slit positions used for spectroscopy.} 
\end{figure}

\clearpage

\begin{figure} 
\figurenum{5} 
\epsscale{0.95} 
\plotone{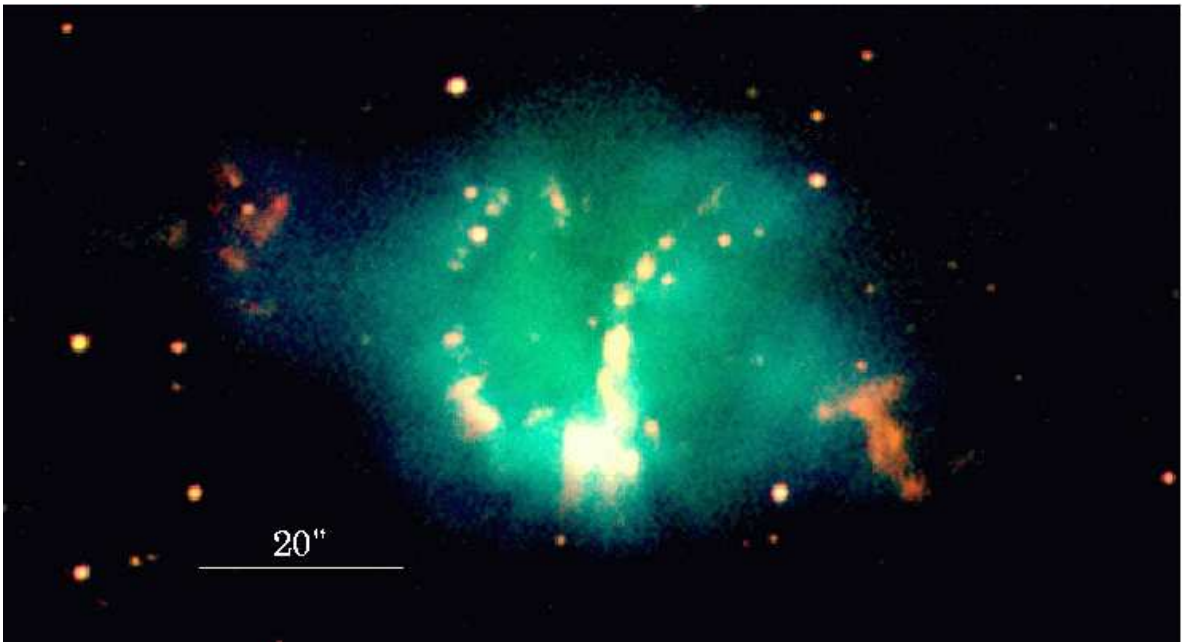} 
\caption{Composite colour picture of PM\,1-333 constructed with the images shown in Fig.\,1. Red 
corresponds to [N\,{\sc ii}], green to H$\alpha$, and blue to [O\,{\sc iii}]. North is up, 
east to the left.} 
\end{figure} 

\clearpage

\begin{figure} 
\figurenum{6} 
\epsscale{0.85} 
\plotone{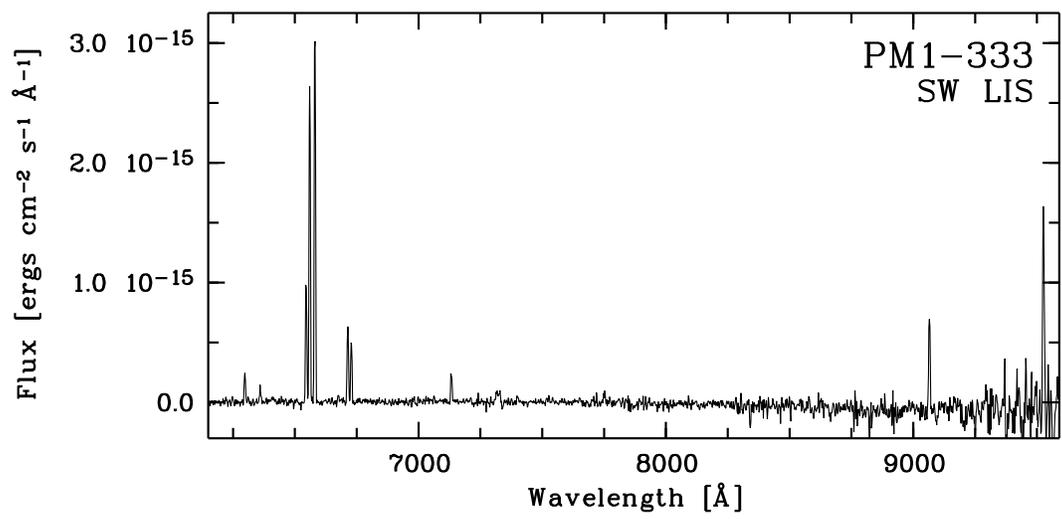} 
\caption{Red 2.2\,m CAHA CAFOS spectrum of the SW low-ionization structure 
of PM\,1-333 (see Fig.\,4).} 
\end{figure}

\clearpage

\begin{deluxetable}{ccccccc}
\tablecaption{Names, coordinates and IRAS colours of PM\,1-242, PM\,1-318 and PM\,1-333}
\tablenum{1}
\tablewidth{0pt}
\tablehead{
\colhead{Object} & \colhead{IRAS name} & \colhead{PN\,G name}  & \colhead{$\alpha$ (J2000)}  &
\colhead{$\delta$ (J2000)} & \colhead{[12]$-$[25]} & \colhead{[25]$-$[60]}
}
\startdata
PM\,1-242 & 18320+0005 & 031.1+03.7 & 18$^{h}$ 34$^{m}$ 38.7$^{s}$ & +00$^{\circ}$ 08$^{'}$ 03$^{''}$ & $\geq$ 1.25 & 0.71 \\
PM\,1-318 & 20077+3722 & 074.5+02.3 & 20$^{h}$ 09$^{m}$ 32.9$^{s}$ & +37$^{\circ}$ 31$^{'}$ 15$^{''}$ & $\geq$ 1.22 & 0.95 \\
PM\,1-333 & 21394+5844 & 100.4+04.6 & 21$^{h}$ 40$^{m}$ 59.1$^{s}$ & +58$^{\circ}$ 58$^{'}$ 37$^{''}$ & $\geq$ 1.66 & 0.07 \\
\enddata

\end{deluxetable}

\clearpage

\begin{deluxetable}{lrccccc}
\tabletypesize{\normalsize}
\setlength{\tabcolsep}{0.07in}
\tablecolumns{6}
\tablewidth{0in}
\tablenum{2}
\tablecaption{Emission line intensities in PM\,1-242, PM\,1-318 and PM\,1-333}
\tablehead{
\colhead{} & \colhead{} & \colhead{PM\,1-242}  & \colhead{PM\,1-318} & \colhead{PM\,1-333} & \colhead{PM\,1-333} & \colhead{PM\,1-333} \\
\colhead{} & \colhead{} & \colhead{}  & \colhead{} & \colhead{(North)} & \colhead{(South)} & \colhead{(SW-LIS)} \\
\colhead{Line} & \colhead{f($\lambda$)} & \colhead{I($\lambda$)} & \colhead{I($\lambda$)} & \colhead{I($\lambda$)} & \colhead{I($\lambda$)} & \colhead{I($\lambda$)} 
}

\startdata
[O\,{\sc ii}]\,$\lambda$3727            & 0.266   & \nodata         & \nodata      & \nodata        & 165$\pm$3      & \nodata  \\
\[[Ne\,{\sc iii}]\,$\lambda$3869        & 0.228   & 86$\pm$20       & \nodata      & 72$\pm$7       & 123.3$\pm$2.1  & \nodata  \\
He\,{\sc i}\,+\,H8\,$\lambda$3889       & 0.223   & \nodata         & \nodata      & \nodata        & 17.7$\pm$1.4   & \nodata   \\
\[[Ne\,{\sc iii}]\,+\,H$\epsilon$\,$\lambda$3968  & 0.203   & \nodata  & \nodata   & \nodata        & 56.2$\pm$1.4   & \nodata   \\
H$\delta$\,+\,He\,{\sc ii}\,$\lambda$4101 & 0.172 & \nodata         & \nodata      & 32$\pm$4       & 27.0$\pm$1.0   & \nodata  \\
H$\gamma$\,$\lambda$4340                & 0.129   & \nodata         & 41$\pm$5     & 44.3$\pm$2.3   & 49.0$\pm$0.9   & \nodata    \\
\[[O\,{\sc iii}]\,$\lambda$4363         & 0.124   & \nodata         & \nodata      & \nodata        & 23.9$\pm$0.7   & \nodata  \\
He\,{\sc ii}\,$\lambda$4686             & 0.042   & 44$\pm$8        & 105$\pm$5    & 92.5$\pm$2.1   & 60.8$\pm$0.7   & \nodata   \\
He\,{\sc i}\,+\,\[[Ar\,{\sc iv}]\,$\lambda$4711   & 0.036 & \nodata & \nodata      & 27.0$\pm$1.9   & 11.2$\pm$0.4   & \nodata  \\
\[[Ar\,{\sc iv}]\,$\lambda$4740         & 0.029   & \nodata         & \nodata      & 19.4$\pm$2.1   & 7.2$\pm$0.4    & \nodata    \\
H$\beta$\,$\lambda$4861                 & 0.000   & 100$\pm$6       & 100$\pm$6    & 100.0$\pm$2.0  & 100.0$\pm$0.8  & 100.0$\pm$1.2 \\
\[[O\,{\sc iii}]\,$\lambda$4959         & -0.023  & 473$\pm$9       & 183$\pm$6    & 227$\pm$3      & 398.2$\pm$1.4  & \nodata   \\
\[[O\,{\sc iii}]\,$\lambda$5007         & -0.034  & 1490$\pm$14     & 548$\pm$8    & 642$\pm$4      & 1189.3$\pm$2.3 & \nodata    \\
He\,{\sc ii}\,$\lambda$5411             & -0.118  & \nodata         & 21$\pm$4     & 10.7$\pm$1.2   & 6.6$\pm$0.3    & \nodata    \\
\[[N\,{\sc ii}]\,$\lambda$5755          & -0.191  & \nodata         & \nodata      & \nodata        & 4.4$\pm$0.2    & \nodata   \\
He\,{\sc i}\,$\lambda$5876              & -0.216  & 36$\pm$3        & \nodata      & \nodata        & 6.8$\pm$0.2    & \nodata   \\
\[[O\,{\sc i}]\,$\lambda$6300           & -0.285  & 5.4$\pm$0.8     & \nodata      & \nodata        & 10.8$\pm$0.1   & 28.9$\pm$1.6 \\
\[[S\,{\sc iii}]\,$\lambda$6312         & -0.287  & 5.8$\pm$0.8     & \nodata      & \nodata        & 4.6$\pm$0.1    & 4.2$\pm$0.6   \\
\[[O\,{\sc i}]\,$\lambda$6363           & -0.294  & 2.6$\pm$0.4     & \nodata      & \nodata        & 3.7$\pm$0.1    & 14.6$\pm$1.4   \\
\[[N\,{\sc ii}]\,$\lambda$6548          & -0.321  & 29.8$\pm$0.6    & 1.3$\pm$0.4  & 4.4$\pm$0.4    & 36.8$\pm$0.2   & 110.6$\pm$1.8 \\
H$\alpha$\,$\lambda$6563                & -0.323  & 285.0$\pm$1.4   & 285.0$\pm$1.6 & 285.0$\pm$1.4  & 285.0$\pm$0.5 & 285.0$\pm$2.3 \\
\[[N\,{\sc ii}]\,$\lambda$6584          & -0.326  & 89.3$\pm$0.8    & 3.1$\pm$0.3  & 15.8$\pm$0.5   & 112.1$\pm$0.3  & 323.0$\pm$2.3  \\
He\,{\sc i}\,$\lambda$6678              & -0.338  & 4.0$\pm$0.3     & \nodata      & \nodata        & 3.1$\pm$0.1    & 5.8$\pm$1.5 \\
\[[S\,{\sc ii}]\,$\lambda$6716          & -0.343  & 8.9$\pm$0.4     & \nodata      & \nodata        & 22.7$\pm$0.2   & 68.0$\pm$1.6 \\  
\[[S\,{\sc ii}]\,$\lambda$6731          & -0.345  & 12.8$\pm$0.4    & \nodata      & \nodata        & 20.7$\pm$0.2   & 49.7$\pm$1.6  \\ 
\[[Ar\,{\sc v}]\,$\lambda$7005          & -0.376  & \nodata         & 9.3$\pm$0.6  & \nodata        & \nodata        & \nodata    \\
He\,{\sc i}\,$\lambda$7065              & -0.383  & 4.1$\pm$0.3     & \nodata      & \nodata        & 2.1$\pm$0.1    & 3.6$\pm$0.9  \\
\[[Ar\,{\sc iii}]\,$\lambda$7135        & -0.391  & 30.9$\pm$0.5    & 8.7$\pm$0.6  & 12.7$\pm$0.4   & 19.7$\pm$0.8   & 21.7$\pm$1.0   \\
\[[O\,{\sc ii}]\,$\lambda$7319          & -0.410  & 5.2$\pm$0.5     & \nodata      & \nodata        & 4.1$\pm$0.1    & 10.7$\pm$1.4  \\
\[[O\,{\sc ii}]\,$\lambda$7330          & -0.411  & 4.3$\pm$0.5     & \nodata      & \nodata        & 3.4$\pm$0.1    & 9.2$\pm$1.3  \\
\[[Cl\,{\sc iv}]\,$\lambda$7529         & -0.430  & \nodata         & 1.7$\pm$0.3  & \nodata        & \nodata        & \nodata    \\
He\,{\sc ii}\,$\lambda$7592             & -0.436  & \nodata         & 2.9$\pm$0.6  & 3.4$\pm$0.4    & 1.4$\pm$0.1    & \nodata    \\
\[[Ar\,{\sc iii}]\,$\lambda$7751        & -0.451  & 10.1$\pm$0.4    & 3.4$\pm$0.6  & 8.5$\pm$0.5    & 5.4$\pm$0.1    & 6.5$\pm$1.6 \\
\[[Cl\,{\sc iv}]\,$\lambda$8045         & -0.477  & 2.8$\pm$0.3     & 2.8$\pm$0.5  & \nodata        & \nodata        & \nodata    \\
He\,{\sc ii}\,$\lambda$8236             & -0.492  & 3.0$\pm$0.4     & 6.2$\pm$0.7  & \nodata        & \nodata        & \nodata    \\
P15\,$\lambda$8545                      & -0.532  & \nodata         & \nodata      & \nodata        & 0.6$\pm$0.1    & \nodata   \\
P14\,$\lambda$8598                      & -0.540  & 2.1$\pm$0.6     & \nodata      & \nodata        & 1.0$\pm$0.1    & \nodata  \\
P13\,$\lambda$8664                      & -0.550  & 2.5$\pm$0.6     & \nodata      & \nodata        & 1.1$\pm$0.1    & \nodata   \\
P12\,$\lambda$8750                      & -0.562  & 3.9$\pm$0.6     & \nodata      & \nodata        & 0.9$\pm$0.1    & \nodata    \\
P11\,$\lambda$8862                      & -0.578  & 3.8$\pm$0.6     & \nodata      & \nodata        & 1.3$\pm$0.1    & \nodata     \\
P10\,$\lambda$9014                      & -0.599  & 4.2$\pm$0.7     & 4.2$\pm$1.2  & \nodata        & 2.0$\pm$0.1    & \nodata   \\
\[[S\,{\sc iii}]\,$\lambda$9069         & -0.606  & 71.6$\pm$1.0    & 15.6$\pm$1.1 & 16.5$\pm$0.5   & 28.4$\pm$0.2   & 44.0$\pm$2.1 \\
P9\,$\lambda$9228                       & -0.612  & 10.0$\pm$0.7    & 11.4$\pm$1.8 & \nodata        & 2.2$\pm$0.1    & \nodata    \\
\[[S\,{\sc iii}]\,$\lambda$9532         & -0.620  & 267.0$\pm$2.4   & 28$\pm$3     & 28.8$\pm$1.1   & 51$\pm$5       & 100$\pm$7  \\
P8\,$\lambda$9546                       & -0.620  & 14.6$\pm$1.2    & \nodata      & \nodata        & 2.2$\pm$0.1    & \nodata    \\
\tableline
c(H$\beta$)                             &         & 0.90          & 0.13           & 0.97        & 1.19    & 1.19\tablenotemark{a} \\
log\,F$_{H\beta}$ (erg\,cm$^{-2}$\,s$^{-1}$) &   & $-$13.93      & $-$13.75       & $-$14.43   & $-$13.69  & $-$14.59   \\
\enddata
\tablenotetext{a}{Assumed to be identical to the value in PM\,1-333 South (see text)}
\end{deluxetable}

\clearpage

\begin{deluxetable}{lcc}
\tabletypesize{\normalsize}
\tablenum{3}
\tablecaption{Ionic abundances relative to H$^+$ in PM\,1-242 and PM\,1-333 South}
\tablewidth{0pt}
\tablehead{
\colhead{Ion\tablenotemark{a}} & \colhead{PM\,1-242} & \colhead{PM\,1-333 South}
}
\startdata
He$^+$                          & 0.24$\pm$0.02                     & 0.06$\pm$0.01      \\
He$^{+2}$                       & 0.05$\pm$0.01                     & 0.07$\pm$0.01       \\
O$^0$                           & 1.11$\pm$0.37 $\times$10$^{-5}$   & 5.18$\pm$0.23 $\times$10$^{-6}$     \\
O$^+$                           & 7.62$\pm$4.73 $\times$10$^{-5}$   & 1.47$\pm$0.14 $\times$10$^{-5}$    \\
O$^{+2}$                        & 4.71$\pm$1.27 $\times$10$^{-4}$   & 1.20$\pm$0.04 $\times$10$^{-4}$    \\
N$^+$                           & 1.67$\pm$0.35 $\times$10$^{-5}$   & 8.11$\pm$0.26 $\times$10$^{-6}$    \\
Ne$^{+2}$                       & 8.61$\pm$3.50 $\times$10$^{-5}$   & 3.27$\pm$0.16 $\times$10$^{-5}$   \\
S$^+$                           & 6.66$\pm$1.89 $\times$10$^{-7}$   & 4.44$\pm$0.17 $\times$10$^{-7}$  \\
S$^{+2}$                        & 1.27$\pm$0.20 $\times$10$^{-5}$   & 1.97$\pm$0.11 $\times$10$^{-6}$    \\
Ar$^{+2}$                       & 2.92$\pm$0.53 $\times$10$^{-6}$   & 7.92$\pm$0.52 $\times$10$^{-7}$   \\
Ar$^{+3}$                       & \nodata                           & 5.73$\pm$0.36 $\times$10$^{-7}$   \\
Cl$^{+3}$                       & 1.94$\pm$0.40 $\times$10$^{-7}$   & \nodata       \\
\enddata
\tablenotetext{a}{For ions with more than one transition, a flux-weighted average was obtained}
\end{deluxetable}

\clearpage

\begin{deluxetable}{lcccccc}
\tablenum{4}
\tablecaption{Elemental abundances in PM\,1-242, PM\,1-333 South and other objects}
\tablewidth{0pt}
\tablehead{
\colhead{Element ratio} & \colhead{PM\,1-242\tablenotemark{a}} & \colhead{PM\,1-333 (S)\tablenotemark{a}} 
& \colhead{Type\,I PNe\tablenotemark{b}} & \colhead{Type\,II PNe\tablenotemark{c}} 
& \colhead{H\,{\sc ii} regions\tablenotemark{d}} & \colhead{Sun\tablenotemark{e}}
}
\startdata
He/H                     &  0.29  & 0.13  & 0.13 & 0.11 & 0.10 & 0.09 \\
O/H\,($\times$10$^{4}$)  &  6.21  & 2.25  & 4.47 & 5.13 & 5.01 & 4.57 \\
N/H\,($\times$10$^{4}$)  &  1.36  & 1.24  & 5.25 & 1.58 & 3.71 & 6.02 \\
S/H\,($\times$10$^{5}$)  &  4.66  & 1.05  & 0.81 & 0.60 & 1.15 & 1.38 \\
Ar/H\,($\times$10$^{6}$) &  5.46  & 1.36  & 2.63 & 2.54 & 2.63 & 1.51 \\
Ne/H\,($\times$10$^{4}$) &  1.13  & 0.61  & 1.23 & 1.17 & 0.79 & 0.69 \\
\tableline
N/O                      &  0.22  & 0.55  & 1.17 & 0.31 & 0.07 & 0.13 \\
\enddata
\tablenotetext{a} {See \S3.1 and \S3.3.2 for a discussion on the errors}
\tablenotetext{b} {Average for Type\,I PNe (Kingsburgh \& Barlow 1994)}
\tablenotetext{c} {Average for Type\,II PNe (Henry et al. 2004)}
\tablenotetext{d} {Average for H\,{\sc ii} regions (Shaver et al. 1983)}
\tablenotetext{e} {The Sun (Grevesse et al. 2007)}
\end{deluxetable}

\end{document}